\documentclass[a4paper,11pt]{article}
\pdfoutput=1 

\usepackage{amssymb}
\usepackage{jheppub}
\usepackage{physics}
\usepackage{autobreak}
\usepackage{hyperref}
\usepackage{color}
\usepackage{natbib}
\usepackage{graphicx}
\hypersetup{colorlinks=true, citecolor=blue, urlcolor=blue, linkcolor=blue}

\title{Landau theory, effective temperature, and tricritical phenomena in a holographic nonequilibrium steady state}
\author[a]{Motohiro Kanazawa}
\author[a,b]{Masataka Matsumoto}
\author[a]{Shin Nakamura}
\affiliation[a]{Department of Physics, Chuo University, 1-13-27 Kasuga, Bunkyo-ku, Tokyo 112-8551, Japan}
\affiliation[b]{Department of Physics, Shanghai University, 99 Shangda Road, Shanghai 200444, China}

\abstract{
We investigate the critical exponents $(\gamma,\nu)$ at the current-driven tricritical point (TCP) associated with chiral symmetry breaking in a nonequilibrium steady state described by the D3/D7 model. In the symmetry-broken phase, we find that, unlike in the conventional Landau theory, both $\gamma$ and $\nu$ can take values different from those predicted by the Landau theory and depend on the path along which the TCP is approached in the phase diagram. However, when the TCP is approached with the ratio of the effective temperature to the heat bath temperature, $T_{*}/T$, held fixed, the critical exponents agree with those of the Landau theory. These results suggest that the critical phenomena at the current-driven TCP may be described by the Landau theory along the $T_{*}/T$-fixed line, whereas a nontrivial extension of the Landau theory is required for a more general description of the critical phenomena along arbitrary paths.
}

\begin{document} 
\maketitle
\section{Introduction}
Nonequilibrium steady states (NESSs) are a central target for the deeper understanding of macroscopic properties of systems far from equilibrium. Compared to equilibrium, a steady current, such as a heat flow or electric current, is an additional parameter that characterizes a NESS, and clarifying its role as a thermodynamic variable is a central issue. Extensions of thermodynamics to NESSs have been extensively discussed \cite{oono1998steady,sasa2006steady,nakagawa2019global}, nevertheless, the construction of a universal framework comparable to equilibrium thermodynamics remains a challenging problem.

The AdS/CFT correspondence \cite{Maldacena:1998,Gubser:1998,Witten:1998}, or holography, is a powerful tool to explore strongly correlated quantum systems, relying on a duality in which the strongly coupled dynamics is mapped onto classical gravity in a higher-dimensional spacetime. Owing to this aspect, the holographic approach has been applied to nonequilibrium phenomena, including those in NESS (see reviews \cite{Hubeny:2010ry,Liu:2018crr,Kundu:2019ull}). In particular, phase transitions and critical phenomena in NESS have been studied by using the probe brane model (D3/D7 model), in which a steady current is driven by an external electric field, motivated by exploration of universal properties of critical phenomena in NESS \cite{Nakamura:2012ae,Ali-Akbari:2013hba,Matsumoto:2018ukk}. Interestingly, the electric current naturally arises as a variable that distinguishes the phases of the NESS and plays the role of a macroscopic variable in the critical phenomena. A current-driven phase transition associated with chiral symmetry breaking, as well as a current-driven tricritical point (TCP), has also been found \cite{Imaizumi:2019byu,Matsumoto:2022nqu}.

Since it is closely related to the present work, we briefly summarize the main results of \cite{Matsumoto:2022nqu}. We studied the critical exponents $(\gamma, \nu)$, which characterize the divergence of the susceptibility and the correlation length, respectively, and we found that their values are consistent with those in the Landau theory, that is, the mean-field values, if the TCP is approached from the chiral symmetry-restored phase. 
On the other hand, their values differ from those in the Landau theory if the TCP is approached from the chiral symmetry-broken phase. The asymmetry between the two sides of the transition is a characteristic feature of the current-driven TCP, in contrast to the equilibrium TCP in the same holographic model, where all the critical exponents are consistent with the Landau theory \cite{Matsumoto:2022psr}. Note that critical exponents at a TCP are classified into two categories depending on the direction of approach.
The paths are classified into two types: those tangential to the phase boundary at the TCP and those not. In equilibrium, all non-tangential paths yield a common set of exponents, whereas a tangential approach gives different ones in the Landau theory. In ref.~\cite{Matsumoto:2022nqu}, the TCP is approached along a non-tangential path with fixed dimensionless temperature $T/\sqrt{B}$, where $B$ is the external magnetic field.

A natural question is whether an effective theory, analogous to the Landau theory in equilibrium, exists for the tricritical phenomena in holographic NESSs. In general, holographic models often show the mean-field values of critical exponents since the fluctuations are suppressed in the large $N$ limit. Moreover, the upper critical dimension of an equilibrium TCP  is lowered to $d_{c}=3$ in space, compared with $d_{c}=4$ for an ordinary critical point, so that mean-field exponents are expected in three spatial dimensions (up to logarithmic corrections). However, it is highly nontrivial whether these arguments can be applied to the tricritical phenomena in NESSs. If there exists a Landau theory for the current-driven TCP, identifying the macroscopic variables that constitute such a phenomenological description would provide a crucial hint for understanding universality in NESSs.

Motivated by these questions, we revisit in this paper the tricritical phenomena in the D3/D7 model. In the previous work \cite{Matsumoto:2022nqu}, the phase diagram was drawn in the plane of the dimensionless parameters $(T/\sqrt{B},J/B^{3/2})$, where $J$ is the current density and the magnetic field $B$ is used as normalization, whereas one may use the temperature to introduce dimensionless parameters, $(B/T^{2},J/T^{3})$, as in the earlier study \cite{Imaizumi:2019byu}. 
For our purpose, we employ the latter normalization in this paper.
In systems far from equilibrium, moreover, one can introduce another notion of temperature, the so-called effective temperature (see review \cite{cugliandolo2011effective}). Holographically, it is defined as the Hawking temperature associated with the effective horizon of the open string metric felt by the fluctuations on the probe brane \cite{Kim:2011qh,Sonner:2012if,Nakamura:2013yqa}. We then study the role of the effective temperature $T_{*}$ in the critical phenomena at the TCP, by examining the values of the critical exponents when the TCP is approached from the path with $T_{*}/T$ fixed.\footnote{It has been shown that a nonequilibrium phase transition is well-described in terms of the proper effective temperature in the dragged D5-brane system in \cite{Nakamura:2025grh}.} 

Along the fixed $B/T^{2}$ line, we confirm that $(\gamma,\nu)$ reproduce the values obtained in the previous study \cite{Matsumoto:2022nqu}, which deviate from those in the Landau theory. In contrast, they take the corresponding values in the Landau theory when we approach the TCP 
along the path with $T_{*}/T$ held fixed.
Since $T_{*}$ is the temperature felt by the fluctuations on the probe brane, and $(\gamma,\nu)$ are computed from the two-point function of the fluctuation of the order parameter, holding $T_{*}/T$ fixed is a natural prescription for the tricritical scaling in the NESS. We further find that the scaling relation $\gamma = \nu(2-\eta)$ with $\eta =0$ holds along every path we examine. On the other hand, the exponents are not common for all non-tangential paths (appendix \ref{sec:appA}), suggesting that the classification of paths at an equilibrium TCP does not straightforwardly apply to the NESS.

This paper is organized as follows.
In section \ref{sec:Landau}, we briefly review the critical phenomena at TCP in the equilibrium Landau theory and show the mean-field values of critical exponents.
In section \ref{sec:setup}, we explain our holographic setup based on the D3/D7 model. We outline the background solutions corresponding to the states with the chiral symmetry-broken or restored. We also consider the fluctuations on them to compute the correlation function and the effective temperature.
In section \ref{sec:result}, we show our numerical results of the critical phenomena at the current-driven TCP.
In section \ref{sec:conclusion}, we present our conclusions and discuss some implications of our work.
In appendix \ref{sec:appA}, we discuss the critical exponents with the quantities normalized by the magnetic field.
In appendix \ref{sec:appB}, the quadratic action and the equations of motion for the fluctuations are explicitly presented.

\section{Review: Landau theory for TCP} \label{sec:Landau}
In this section, we briefly review the Landau theory for the TCP in equilibrium. 
In particular, we discuss the critical exponents $(\gamma,\nu)$ for our purpose.
A detailed discussion of the critical phenomena at the TCP can be found, for example, in \cite{lawrie1984phase}.

In the vicinity of the TCP, we expand the free energy functional in powers of the order parameter $M$ up to a sixth-order term:
	\begin{equation}
		f(M) = f_{0}  + \frac{a}{2}M^{2} + \frac{b} {4}M^{4}+ \frac{c}{6} M^{6} - H M, 
		\label{eq:free}
	\end{equation}
where $H$ is a source of the order parameter and $f_{0}$ is a constant.
We assume that $f_{0}=0$ without loss of generality.
We also assume that $c$ is a positive constant, while $a$ and $b$ can be positive and negative values.
The expectation value of the order parameter satisfies the stationary condition for the energy functional:
	\begin{equation}
		\frac{\partial f}{\partial M}=0 \hspace{0.5em} \Longleftrightarrow \hspace{0.5em} H=aM +bM^{3} + cM^{5}.
		\label{eq:expsigma}
	\end{equation}
Let us consider critical phenomena on the $H=0$ plane.
Then, $M=0$ is always a solution with $f(M=0) = 0$.
Now the question is whether other solutions with $f<0$ are possible.
Solving \eqref{eq:expsigma} with $H=0$, we obtain
\begin{equation}
	M^{2} =  \frac{-b \pm \sqrt{b^{2}-4ac}}{2c}.
	\label{eq:OP1}
\end{equation}

For $a>0$ and $b>0$, the only real solution is $M=0$ because $\sqrt{b^{2}-4ac}<b$.
This region corresponds to the symmetry-restored phase.
For $a<0$ and $b>0$, on the other hand, $M=0$ becomes a local maximum of $f$, and we obtain two solutions corresponding to global minima:
\begin{equation}
	M_{\pm} =  \pm \left( \frac{-b + \sqrt{b^{2}-4ac}}{2c} \right)^{1/2},
	\label{eq:OP2}
\end{equation}
where the subscript $\pm$ denotes the sign of the right-hand side. Since $f(M_{\pm})<f(0)=0$, this region corresponds to the symmetry-broken phase.
Note that we exclude the solution with the minus sign in \eqref{eq:OP1}, as it does not yield a real solution.
Since the solutions $M=0$ and $M=M_{\pm}$ become degenerate in the limit of $a\to 0$, the line $a=a_{\lambda}(b)=0$ can be identified as a line of critical points, the so-called $\lambda$-line.

For $b<0$, there are two specific values of $a$. One is $a=b^{2}/(4c)$ at which the term under the square root vanishes.
For $a>b^{2}/(4c)$, the only real solution is $M=0$ again.
For $a<b^{2}/(4c)$, on the other hand, the solutions $M_{\pm}$ corresponding to local minima and the solutions
\begin{equation}
    \tilde{M}_{\pm} =  \pm \left( \frac{-b - \sqrt{b^{2}-4ac}}{2c} \right)^{1/2},
	\label{eq:OP3}
\end{equation}
corresponding to local maxima appear.
The other is the value at which the three phases coexist:
	\begin{equation}
		f(M_{\pm}) = f(0) \hspace{0.5em} \Longleftrightarrow \hspace{0.5em} a=\frac{3b^{2}}{16 c}. \label{eq:tau-line}
	\end{equation}

In other words, the line $a=a_{\tau}(b) = {3b^{2}}/{(16 c)}$ is the three-phase coexistence line, along which a first-order phase transition occurs. This line is known as the $\tau$-line. 
Since $f(M_{\pm})<0$ below the $\tau$-line $(a<a_{\tau}(b))$, this region corresponds to the symmetry-broken phase.
The TCP is defined by the endpoint of the $\tau$-line, namely $a=b=0$.
	
For our purpose, we consider the susceptibility defined as
	\begin{equation}
		\chi \equiv \left. \frac{\partial M}{\partial H} \right|_{H=0}= \frac{1}{a+3bM^{2}+5cM^{4}},
	\end{equation}
where we obtain this by differentiating (\ref{eq:expsigma}) with respect to $H$. 
Here, $M$ is the solution given by (\ref{eq:OP2}), and we omit the subscript $\pm$ for simplicity.

We now consider several paths to the TCP to determine the critical exponent $\gamma$, which characterizes the divergence of the susceptibility.
Firstly, we consider paths along $b=0$ with $a\to 0^{\pm}$, which correspond to the path in the symmetry-restored phase and symmetry-broken phase, respectively.
The susceptibility along this line is given by
	\begin{equation}
		\chi =\frac{1}{a+5c M^{4}}=
        \begin{cases}
            1/a & \text{for $M=0$} \\
            -1/(4a) & \text{for $M^{2}= \sqrt{-a/c}$}
        \end{cases},
		\label{eq:chi0}
	\end{equation}
where $M=0$ for $a>0$ and $M^{2}=\sqrt{-a/c}$ for $a<0$.
For both cases, we obtain 
	\begin{equation}
		\chi \propto \abs{a}^{-\gamma},
		\label{eq:chi2}
	\end{equation}
with $\gamma=1$. 

Secondly, we consider general paths defined by $a p + b q =0$ with arbitrary constants $p$ and $q$.
As we have already considered the path with $b=0$ above, we now assume that $p>0$ without loss of generality. Then, the order parameter is determined by the solution to
\begin{equation}
    -\frac{q}{p}b M+ bM^{3} +c M^{5} =0, \label{eq:Mg}
\end{equation}
and the susceptibility is
\begin{equation}
    \chi = \frac{1}{\left(-\frac{q}{p} + 3M^{2} \right)b + 5c M^{4}}.
\end{equation}
In the vicinity of the TCP, the susceptibility along this general line is then given by
\begin{equation}
    \chi = 
    \begin{cases}
        -p/(qb) & \text{for $M=0$} \\
        p/(4qb) +\cdots & \text{for $M = M_{g}$} 
    \end{cases},
\end{equation}
where the ellipsis denotes subleading terms in the limit of $b\to 0$ and $M_{g}$ is the solution to \eqref{eq:Mg}.
Therefore, we obtain
\begin{equation}
    \chi \propto \abs{b} ^{-\gamma},
\end{equation}
with $\gamma =1 $ again.

Lastly, we consider a path along the $\tau$-line, defined by \eqref{eq:tau-line}. In this case, the order parameter is determined by the solution to 
\begin{equation}
    \frac{3b^{2}}{16 c} M + b M^{3} + c M^{5} =0,
\end{equation}
and the susceptibility is 
\begin{equation}
    \chi = \frac{1}{\frac{3b^{2}}{16 c}+ 3 b M^{2}  + 5c M^{4}}.
\end{equation}
The nontrivial solutions are explicitly given by
\begin{equation}
    M^{2}=-\frac{b}{2c} \pm \frac{b}{4c},
\end{equation}
which correspond to the local maxima and global minima, respectively.
Thus, the susceptibility evaluated at the global minima becomes
\begin{equation}
    \chi = \frac{4 c}{3 b^{2}}.
\end{equation}
As a result, as $b\to 0$ along the $\tau$-line, we find
\begin{equation}
    \chi \propto (-b)^{-\gamma},
\end{equation}
with $\gamma =2$.

To study the behavior of the correlation function and determine the critical exponent $\nu$, we consider the spatial dependence of the order parameter.
Here, the order parameter is given by the expectation value of the microscopic fluctuating variable, $\phi(x)$:
	\begin{equation}
		M(x)=\expval{\phi(x)}.
	\end{equation}
If we assume that the source $H$ also depends on the spatial coordinate, the free energy functional is given by
	\begin{equation}
		f(\phi) = \int d^{3}x \left( -H(x)\phi(x) + \frac{1}{2}\left[ \nabla \phi(x) \right]^{2} + V\left[\phi(x) \right] \right),
	\end{equation}
where 
	\begin{equation}
		V\left[\phi(x) \right]\equiv \frac{a}{2}\phi(x)^{2} + \frac{b}{4}\phi(x)^{4} + \frac{c}{6}\phi(x)^{6},
	\end{equation}
and we ignore the constant term $f_{0}$.
If the source is spatially uniform, the local minimum of $f$ gives a spatially uniform order parameter $\bar{M}$ because the gradient term in $f$ is non-negative.
The equation of state is given by
	\begin{equation}
		H(x) = -\nabla^{2} M(x) +V'\left[ M(x)\right].
	\end{equation}
Taking the derivative of this with respect to $H(x^{\prime})$, we obtain 
	\begin{equation}
		\delta(x-x^{\prime}) = \left( -\nabla^{2} +V''\left(M \right) \right) G(x-x^{\prime}),
	\end{equation}
where
	\begin{equation}
		G(x-x^{\prime}) = \expval{\left(\phi(x)- \bar{M}\right)\left(\phi(x^{\prime})- \bar{M}\right)} = \frac{\delta M(x)}{\delta H(x^{\prime})},
	\end{equation}
which is the correlation function of the order parameter.
Here, $H$ and $M$ are taken to be spatially uniform after differentiation.
The correlation function is given by the Ornstein-Zernike form (with $x^{\prime}=0$):
	\begin{equation}
		G(x) = \frac{1}{(2\pi)^{3}}\int d^{3}k \,e^{i k x} \frac{1}{k^{2}+\chi^{-1}},
	\end{equation}
where $\chi = V''\left[M \right]^{-1}$ is the susceptibility.
When $\abs{x}$ is large, the correlation function behaves as $G(x) \sim \exp\left( -\abs{x} / \sqrt{ \chi} \right)$, and the correlation length is given by $\xi = \sqrt{\chi}$.
In the Landau theory, thus, the divergence of the correlation length near the TCP is related to that of the susceptibility:
	\begin{equation}
		\xi \propto \abs{a}^{-\nu} \sim \abs{a}^{-\gamma/2}.
		\label{eq:nu}
	\end{equation}
The same relation holds when $\chi$ is expressed in terms of $|b|$.
In addition, the correlation function of the order parameter in Fourier space is given by
	\begin{equation}
		\tilde{G}(k) = \frac{1}{k^{2}+\chi^{-1}}.
	\end{equation}
Since the susceptibility diverges at the TCP, we find $ \tilde{G}(k)|_{\rm TCP}\sim k^{\eta-2}$ with $\eta=0$.
Note that considering the scaling hypothesis of the correlation function, one can generally derive the scaling relation among these critical exponents:
\begin{equation}
    \gamma = \nu (2-\eta). \label{eq:scaling}
\end{equation}

In summary, the critical exponents $(\gamma,\nu)$ at the TCP are $(1,1/2)$ or $(2,1)$ in the Landau theory.
Since fluctuations of the order parameter are generally suppressed in Large-$N$ gauge theories, the critical exponents are expected to agree with the mean-field values (the values in the Landau theory).
Indeed, most of the existing holographic calculations of critical exponents yield values consistent with the Landau theory
with a few counterexamples \cite{Matsumoto:2022nqu} (see, e.g. \cite{Natsuume:2010vb} for another example).
Moreover, the upper critical dimension of the TCP is known to be $d_{c}=3$ \cite{lawrie1984phase}.
Since the present system has three spatial dimensions, the critical exponents which agree with the values in the Landau theory are expected from this viewpoint as well.
Nevertheless, as we show in the following sections, we find that the values of $(\gamma,\nu)$ at the current-driven TCP studied here do not agree with those of the Landau theory, even though we are dealing with a Large-$N$ gauge theory with three spatial dimensions.

\section{Holographic setup} \label{sec:setup}
In this section, we briefly review our holographic model, which exhibits spontaneous chiral symmetry breaking and gives rise to the current-driven nonequilibrium TCP in the phase diagram \cite{Imaizumi:2019byu,Matsumoto:2022nqu}. We also consider fluctuations around the background solutions in both the chiral-symmetry-broken and chiral-symmetry-restored phases to compute the critical exponents $(\gamma,\nu)$ associated with the correlation function. We then introduce the effective temperature, defined as the Hawking temperature associated with the effective horizon that emerges on the D7-brane in a NESS.

\subsection{Background solutions}
We consider strongly coupled $SU(N)$ $\mathcal{N}=4$ supersymmetric Yang--Mills (SYM) theory with $\mathcal{N}=2$ hypermultiplet in the large-$N$ limit. The charged particles, which carry a global $U(1)$ charge, belong to the hypermultiplet sector, whereas the degrees of freedom forming the heat bath belong to the $\mathcal{N}=4$ SYM sector. When an external electric field is applied, the system reaches a NESS with a finite current density. Since the charged particles in our system are massless, the theory has a $U(1)$ chiral symmetry at the level of the Lagrangian. Furthermore, we take the system to be neutral, with equal numbers of positively and negatively charged particles. We also apply an external magnetic field perpendicular to the electric field. In this paper, we study phase transitions in which chiral symmetry is spontaneously broken at finite current density in the presence of the external electromagnetic fields.

The holographic dual of our system is the D3/D7 model \cite{Karch:2002sh} in the presence of external electromagnetic fields \cite{Karch:2007pd,Ammon:2009jt}, treated in the probe limit. The background geometry is the five-dimensional AdS--Schwarzschild black hole times $S^5$:
\begin{equation}
ds^2=\frac{L^2}{u^2}\left(-f(u)dt^2+\frac{du^2}{f(u)}+d\vec{x}^2\right)+L^2d\Omega^2_{5} ,
\end{equation}
where $f(u)=1-u^4/u^4_{\mathrm{H}}$. $u$ is the radial coordinate that ranges from $0$ to $u_{\mathrm{H}}$, while $t$ and $\vec{x}=(x,y,z)$ are the coordinates of the $(3+1)$-dimensional spacetime of the gauge theory. The black hole horizon is located at $u=u_{\mathrm{H}}$, while the boundary is at $u=0$. The Hawking temperature is given by $T=1/(\pi u_{\mathrm{H}})$ and is identified with the temperature of the heat bath on the gauge theory side. The line element on $S^5$ is given by
\begin{equation}
   d\Omega^2_{5}=d\theta^2+\sin^2 \theta d\psi^2 +\cos^2\theta d\Omega^2_{3} ,
\end{equation}
where $d\Omega^2_{3}$ is the line element of the unit $S^3$. The D7-brane wraps the $S^3$ part.

The dynamics of the D7-brane is governed by the Dirac-Born-Infeld (DBI) action
\begin{equation}
    S_{\mathrm{DBI}}=-T_{\mathrm{D7}}\int d^8\xi\sqrt{-\det (g_{ab}+(2\pi l^2_s)F_{ab})} ,
\end{equation}
where $g_{ab}$ is the induced metric and $F_{ab}=\partial_a A_b-\partial_b A_a$ is the field strength of the $U(1)$ gauge field $A_a$ on the D7-brane. The D7-brane tension is given by $T_{\rm D7}^{-1}=(2\pi)^7 l_s^8 g_s$, where $l_s$ and $g_s$ are the string length and string coupling constant, respectively. We set $L=1$ and $2\pi l_s^2=1$ for simplicity. This choice corresponds to setting $2\lambda=(2\pi)^2$, where $\lambda=g_{\rm YM}^2N$ is the 't~Hooft coupling of the gauge theory.

The configuration of the D7-brane is specified by $\theta(u)$ ($0\leq\theta\leq\pi/2$) and $\psi$. We set $\psi=0$ without loss of generality. Since $\theta$ depends only on $u$, the nontrivial component of the induced metric is $g_{uu}=1/\left(u^2 f(u)\right)+\theta^{'}(u)^2$, where the prime denotes differentiation with respect to $u$. In addition, we apply an electric field in the $x$ direction and a magnetic field in the $z$ direction. We employ the following ansatz for the gauge fields:
\begin{equation}
    A_{x}(t,u)=-Et+h(u),\quad A_{y}(x)=Bx , 
\end{equation}
where $E$ and $B$ denote the electric and magnetic fields acting on the charged particles, respectively. With this ansatz, the DBI action becomes
\begin{equation}
    S_{\mathrm{DBI}}=\int d^5\xi{\cal L}_{\rm D7}=-\mathcal{N}\int d^5\xi\sqrt{- \cos^6\theta \left( g_{xx} g_{uu}(g_{tt}g^2_{xx}+g_{tt}B^2+g_{xx}E^2) +g_{tt} g_{xx}^2 (h')^2 \right)},
\end{equation}
where $\mathcal{N}=T_{\mathrm{D7}}(2\pi^2)$.
The expectation value of the current density is given by $J=-\partial{\cal L}_{\rm D7}/\partial h'$, which is a constant independent of $u$.
One obtains
\begin{eqnarray}
(h')^2=-\frac{g_{uu}(g_{tt}g^2_{xx}+g_{tt}B^2+g_{xx}E^2)}{g_{tt}g_{xx}(J^2+\mathcal{N}^2g_{tt}g^2_{xx}\cos^6{\theta})} J^2.  
\label{h-prime-sq}
\end{eqnarray}
Since $h^{\prime}$ must be real, we require that the denominator and the numerator of (\ref{h-prime-sq}) change their sign simultaneously at $u=u_*$:
\begin{align}
    &\left.|g_{tt}|(g^2_{xx}+B^2)-g_{xx}E^2\right|_{u=u_*}=0,\label{eq:ustar}\\
    &\left.\mathcal{N}^2|g_{tt}|g^2_{xx}\cos^6{\theta}-J^2\right|_{u=u_*}=0.\label{eq:J}
\end{align}
The first condition determines the location of $u_*$, while the second condition gives the expression for $J$.
It will be found that $u=u_*$ is the location of the \textit{effective horizon} discussed in section \ref{subsec:effTemp}.

In the vicinity of the boundary, $\theta(u)$ and $h(u)$ can be expanded as follows:
\begin{equation}    \theta(u)=mu+\theta_2u^3+\cdots , \quad h(u)=\frac{J}{2\mathcal{N}}u^2+\cdots , 
\end{equation}
where $m$ is the mass of the charged particles. Note that the current is in the $x$ direction even in the presence of the magnetic field since the total charge density is zero.
The operator conjugate to $m$ is the chiral condensate given by $\langle \bar{q}q \rangle=\mathcal{N}(2\theta_{2}-m^3/6)$ \cite{Karch:2007pd}.
Hereafter, we set $\mathcal{N}=1$ for simplicity.

We solve the equation of motion for $\theta(u)$ with the following boundary conditions. 
At the boundary, we impose the massless condition $m=\left.u^{-1}\theta(u)\right|_{u=0}=0$. 
At the effective horizon, we impose the regularity condition on $\theta(u)$ there.
Setting $u_{\mathrm{H}}=1$, the explicit form of the regularity condition is given by
\begin{equation}
    \theta'(u_*) = \frac{C-\sqrt{C^2-3D}}{u_{*}D},
\end{equation}
where
\begin{equation}
    C=(3-u^4_{*})\cot{\theta(u_{*})} , \quad D=3(u^4_{*}-1).
\end{equation}
$\theta(u_*)$ is a parameter we set in the numerical computations and $u_{*}$ is determined by Eq.~\eqref{eq:ustar}.

\begin{figure}
    \centering
    \includegraphics[width=0.9\linewidth]{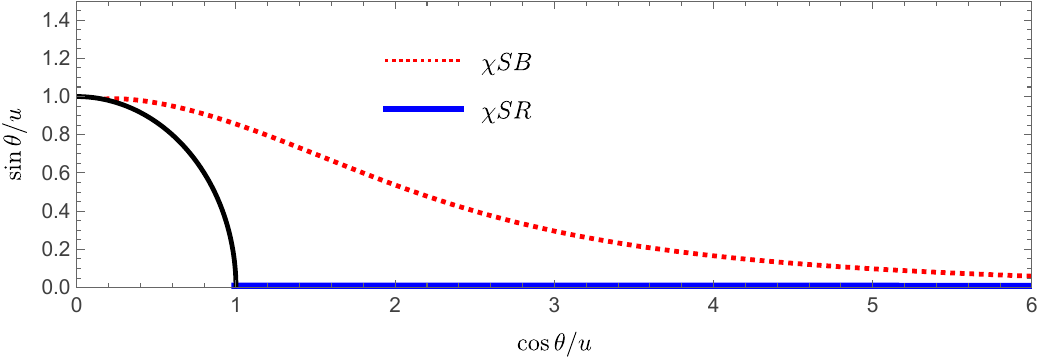}
    \caption{The typical configurations of the D7-brane, corresponding to the $\chi$SB (red dotted) and $\chi$SR (blue solid) phases. The black solid circle corresponds to the location of the effective horizon $u=u_{*}$.}
    \label{fig:config}
\end{figure}

We find two types of background solutions, $\theta(u)=0$ and $\theta(u)\neq0$, distinguished by the chiral condensate $\langle\bar{q}q\rangle$, which serves as the order parameter. The former solution, which shows that the chiral condensate vanishes ($\langle\bar{q}q\rangle=0$), holds regardless of $E$ and $B$.
The latter solution with non-vanishing chiral condensate is possible only in the presence of $B$. 
These two types of solutions correspond to the chiral symmetry-restored ($\chi$SR) phase and the chiral symmetry-broken phase ($\chi$SB), respectively \cite{Babington:2003vm}. In Figure \ref{fig:config}, we show the typical background solutions obtained by numerically solving the equation of motion for $\theta(u)$. The D7-brane configuration for the $\chi$SR  phase corresponds to the trivial solution with $\theta(u)=0$, whereas for the $\chi$SB phase a nontrivial one with $\theta(u)\neq 0$. Note that both solutions asymptotically approach $\theta(u)\to 0$ when $u\to 0$ since we deal with spontaneous chiral symmetry breaking.

\subsection{Fluctuations}
Here we consider the fluctuations on the background solutions $\{\theta,A_{x}\}$ in order to compute the static susceptibility and the correlation length and determine the critical exponents $\{\gamma,\nu\}$. Our ansatz is 
\begin{equation}
    \theta \to \theta(u) + \delta \theta(z,u), \quad A_{x} \to - E t + h(u) + \delta a_{x}(z,u),
\end{equation}
where we assume that the fluctuations depend on a specific spatial direction $z$ to derive the correlation length.
The two fluctuations decouple in the symmetry-restored phase, whereas they are coupled through the nontrivial background configuration $\theta(u)$ in the broken phase.
Note that the fluctuation of $A_{y}$ does not couple to them by choosing the $z$ direction as the spatial dependence. If we expand the action up to the quadratic order in $\tilde{\Phi} =\{\delta \theta, \delta a_{x}\}^{T}$, we obtain the quadratic action
\begin{equation}
	S^{(2)} = -\frac{\mathcal{N}}{2}\int \dd[4]{x}\dd{u} R(u)\Big[
		\partial_{\alpha} \tilde{\Phi}^T \tilde{A}^{\alpha\beta} \partial_{\beta} \tilde{\Phi}
  		+ 2 \tilde{\Phi}^T \tilde{B}^{\alpha} \partial_{\alpha} \tilde{\Phi}
  		+ \tilde{\Phi}^T \tilde{C} \tilde{\Phi}
	\Big],
	\label{eq:effective_action}
\end{equation}
where $R(u) = -{\cal{L}}_{\rm D7}/{\cal{N}}$ and the coefficient matrices, $\{\tilde{A}^{\alpha\beta},\tilde{B}^{\alpha}, \tilde{C} \}$, are explicitly given in appendix \ref{sec:appB}.
Performing the Fourier transformation
\begin{equation}
    \delta \theta(z,u) = \int \dd k\, \vartheta(k,u) e^{ i k z}, \quad \delta a_{x}(z,u) = \int \dd k \, a(k,u) e^{i k z},
 \end{equation}
we derive the coupled differential equations for $\{ \vartheta, a \}$, whose explicit forms are shown in appendix \ref{sec:appB} because they are quite lengthy and complicated.
Since we are interested in the static correlation of $\vartheta$, we impose the regularity condition at the effective horizon without choosing the ingoing/outgoing wave boundary condition.
The asymptotic behavior of $\vartheta$ is given by
\begin{equation}
    \frac{\vartheta(k,u)}{u}  = \delta m(k) + \delta c(k) \,u^{2} + \cdots,
\end{equation}
and the correlation function is obtained by the holographic prescription:
\begin{equation}
    \tilde{G}(k) \equiv \frac{\delta c(k)}{\delta m (k)} \propto  \frac{\delta^{2} S^{(2)}_{\rm on-shell}}{\delta m(k) \delta m(-k)} , \label{eq:corr1}
\end{equation}
where $S^{(2)}_{\rm on-shell}$ is the renormalized on-shell action and $\delta m$ is the source for the operator dual to $\vartheta$. 
Thus, the homogeneous susceptibility corresponds to $\chi = \tilde{G}(k=0)$ and the correlation length is read off from the pole of $\tilde{G}(k)$ in the complex $k$-plane, assuming the Ornstein--Zernike form:
\begin{equation}
    \tilde{G}(k) \sim \frac{1}{k^{2} + 1/\xi^{2}}, \label{eq:corr2}
\end{equation}
where $\xi$ is the correlation length, which diverges at the critical point. Note that we also assume that $\eta = 0$ because $\tilde{G} \sim k^{-2}$ at the critical point. This form is a good approximation for $k\xi \ll 1$, and its validity in the present system was verified in the previous work \cite{Matsumoto:2022nqu}.

\subsection{Effective temperature}
\label{subsec:effTemp}
When world-volume gauge fields on the D7-brane are turned on, the open strings on the D7-brane are governed by the open-string metric (OSM) \cite{Kim:2011qh,Sonner:2012if,Nakamura:2013yqa}
\begin{equation}
    G_{ab}=g_{ab}-(Fg^{-1}F)_{ab}. \label{eq:osm}
\end{equation}
Note that we have set $(2\pi l_s^2)=1$.
The non-vanishing components of the field strength are $F_{tx}=-F_{xt}=-E$, $F_{xy}=-F_{yx}=B$ and $F_{xu}=-F_{ux}=-h'(u)$ in our ansatz.
Substituting these into $G_{ab}$ generates off-diagonal components as follows:
\begin{equation}
    G_{ty}=G_{yt}=\frac{EB}{g_{xx}}, \quad G_{tu}=G_{ut}=-\frac{Eh'(u)}{g_{xx}}, \quad G_{yu}=G_{uy}=-\frac{Bh'(u)}{g_{xx}},
\end{equation}
so that $g_{ab}$ and $G_{ab}$ have different causal structures. In order to diagonalize the OSM, we consider the following transformation for $t$, $y$ and $u$ \cite{Hoshino:2014nfa}:
\begin{equation}
    \begin{pmatrix} dt \\ dy \\ du \end{pmatrix}
    \longrightarrow
    \begin{pmatrix} d\tau \\ d\eta \\ d\rho \end{pmatrix}
    =
    \begin{pmatrix}
        dt + \frac{G_{ty}G_{yu}-G_{tu}G_{yy}}{(G_{ty})^2-G_{tt}G_{yy}}\,du \\
        dy + \frac{G_{ty}}{G_{yy}}\,dt + \frac{G_{yu}}{G_{yy}}\,du \\
        du
    \end{pmatrix} ,
\end{equation}
and then the diagonalized metric $\tilde{G}_{ab}$ (which we call the effective metric) is
\begin{equation}
\begin{aligned}
    \tilde{G}_{\tau\tau} &= -\frac{|g_{tt}|(g^2_{xx}+B^2)-g_{xx}E^2}{B^2+g^2_{xx}}, &
    \tilde{G}_{xx} &= \frac{g_{xx}\cos^6\theta(u)\left(|g_{tt}|(g_{xx}^2+B^2) - g_{xx}E^2\right)}{|g_{tt}|g_{xx}^2\cos^6\theta(u)-J^2}, \\
    \tilde{G}_{\eta\eta} &= \frac{B^2+g_{xx}^2}{g_{xx}}, &
    \tilde{G}_{zz} &= g_{xx}, \\
    \tilde{G}_{\rho\rho} &= \frac{|g_{tt}|g^2_{xx}g_{uu}\cos^6{\theta(u)}}{|g_{tt}|g^2_{xx}\cos^6{\theta(u)}-J^2}, &
    \tilde{G}_{\Omega_3\Omega_3} &= g_{\Omega_3\Omega_3}.
\end{aligned}
\end{equation}
The numerator of $\tilde{G}_{\tau\tau}$ vanishes at $u=u_*$, as can be seen from Eq.~\eqref{eq:ustar}, while the denominator remains finite there. 
Similarly, the denominator of $\tilde{G}_{\rho\rho}$ goes to zero at the same point, as shown in Eq.~\eqref{eq:J}, whereas its numerator stays finite.
Note that $\tilde{G}_{xx}$ remains finite at $u=u_*$, although the numerator and the denominator vanish there.
Hence, near $u_*$, the $\tau\tau$-component and $\rho\rho$-component of the effective metric behave as follows:
\begin{align}
    \tilde{G}_{\tau\tau}&\sim -\left.\frac{(|g_{tt}|(g^2_{xx}+B^2)-g_{xx}E^2)'}{B^2+g^2_{xx}}\right|_{u=u_*}(u-u_*),\\
     \tilde{G}_{\rho\rho}&\sim \left.\frac{|g_{tt}|g^2_{xx}g_{uu}\cos^6{\theta(u)}}{(|g_{tt}|g^2_{xx}\cos^6{\theta(u)})'}\right|_{u=u_*}
     \frac{1}{(u-u_*)}.
\end{align}
This is precisely the signature of a horizon: the effective horizon.
The effective horizon gives the causal boundary for the fluctuations governed by the OSM.
Then, the Hawking temperature associated with the effective horizon, which we call effective temperature $T_{*}$, is given by
\begin{equation}
    T_{*}=\left.\frac{1}{4\pi}\sqrt{\frac{(|g_{tt}|(g^2_{xx}+B^2)-g_{xx}E^2)'(|g_{tt}|g^2_{xx}\cos^6{\theta(u)})'}{(B^2+g^2_{xx})(|g_{tt}|g^2_{xx}g_{uu}\cos^6{\theta(u)})}}\right|_{u=u_{*}}. \label{eq:Teff}
\end{equation}

\section{Results} \label{sec:result}
\begin{figure}
    \centering
    \includegraphics[width=0.85\linewidth]{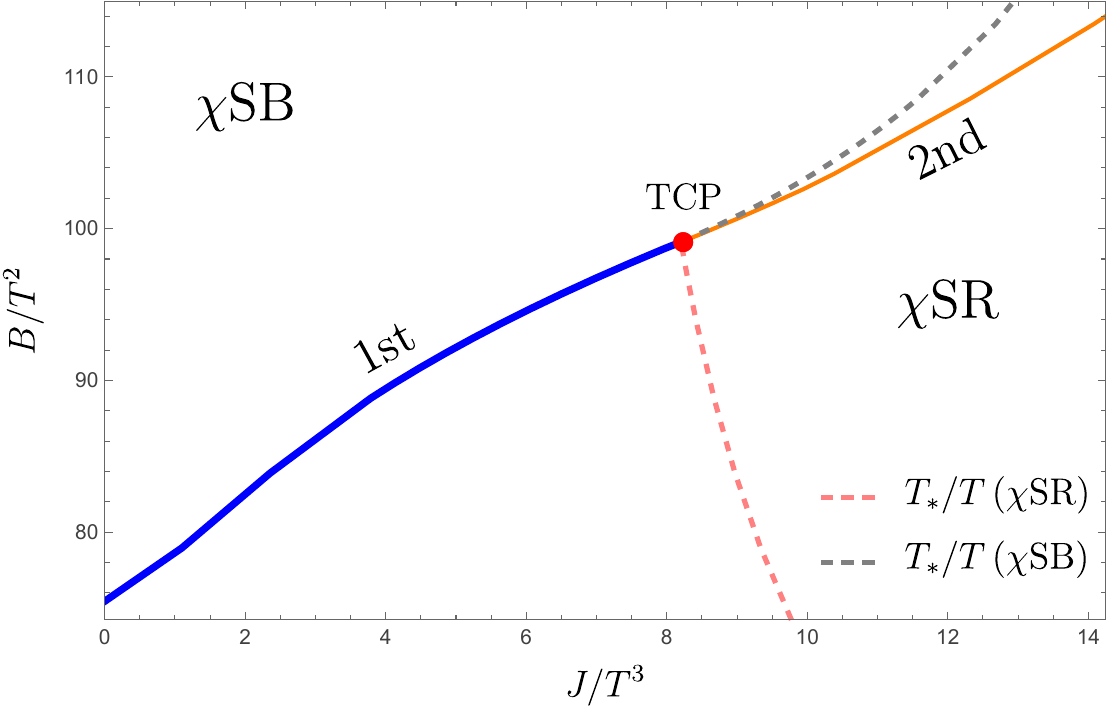}
    \caption{Phase diagram with the plane of $(B/T^{2},J/T^{3})$, originally shown in \cite{Imaizumi:2019byu}. The chiral symmetry-broken ($\chi$SB) phase and chiral symmetry-restored ($\chi$SR)  phase are separated by the first-order phase transition line (blue solid) and second-order phase transition line (orange solid). 
    The point where the first-order phase transition line changes into the second-order phase transition line is the TCP (red circle).
    The line with constant $T_{*}/T$ that goes across the TCP is drawn by the dashed line.}
    \label{fig:phase1}
\end{figure}

In this section, we discuss the numerical results of the critical phenomena at the current-driven TCP.
First of all, we reproduce the phase diagram drawn in the plane of $(B/T^{2},J/T^{3})$ \cite{Imaizumi:2019byu} in figure \ref{fig:phase1}. The $\chi$SB phase and $\chi$SR phase are separated by the first-order and second-order phase transition lines. The endpoint of the first-order and second-order phase transition lines corresponds to the TCP. For later convenience, we also draw the line of fixed $T_{*}/T$ for both phases.

\begin{figure}
    \centering
    \includegraphics[width=0.95\linewidth]{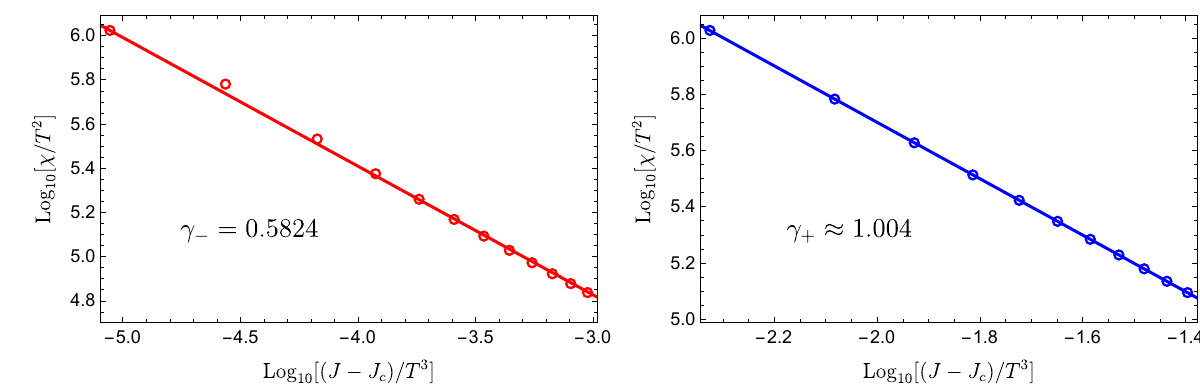}
    \caption{The susceptibility $\chi/T^{2}$ as a function of $(J-J_{c})/T^{3}$ on a logarithmic scale when the TCP is approached from the $\chi$SB phase (left) and $\chi$SR phase (right) along the fixed $B/T^{2}$ line. The data points are the numerical results and the solid line denotes the linear fit to them. The corresponding values of $\gamma$ are also shown. }
    \label{fig:gammaB}
\end{figure}

\begin{figure}
    \centering
    \includegraphics[width=0.95\linewidth]{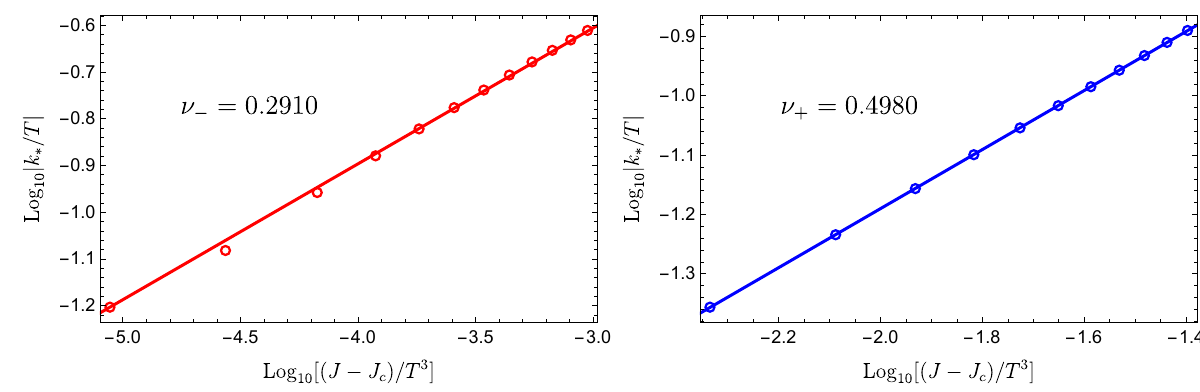}
    \caption{The inverse of the correlation length $k_{*}/T$ as a function of $(J-J_{c})/T^{3}$ in logarithmic scale when the TCP is approached from the $\chi$SB phase (left) and $\chi$SR phase (right) along the fixed $B/T^{2}$ line. The data points are the numerical results and the solid line denotes the linear fit to them. The corresponding values of $\nu$ are also shown.}
    \label{fig:nuB}
\end{figure}

We now study the critical phenomena for $(\gamma,\nu)$ at the TCP. Following the previous works \cite{Imaizumi:2019byu,Matsumoto:2022nqu}, we define these critical exponents as
\begin{equation}
    \tilde{\chi} \propto \abs{\tilde{J}-\tilde{J}_{c}}^{-\gamma}, \quad \tilde{\xi} \propto \abs{\tilde{J}-\tilde{J}_{c}}^{-\nu},
\end{equation}
where the quantities with tilde are normalized by $T$, such as $\tilde{\chi}=\chi /T^{2}$, $\tilde{\xi} = \xi T$, and $\tilde{J} = J/T^{3}$ with the homogeneous susceptibility $\chi$ and the correlation length $\xi$, introduced in \eqref{eq:corr1} and \eqref{eq:corr2}. 
We first consider the path of the fixed $B/T^{2}$ with the value of that at the TCP from both the $\chi$SB phase and the $\chi$SR phase. In figure \ref{fig:gammaB}, we show the susceptibility as a function of $(\tilde{J}-\tilde{J}_{c})$ on a logarithmic scale as approached from the $\chi$SB phase (left) and $\chi$SR phase (right). From the linear fitting, we obtain
\begin{equation}
    \gamma_{-} \approx 0.5824, \quad \gamma_{+} \approx 1.006,
\end{equation}
where $\mp$ subscripts indicate the values of the $\chi$SB phase and $\chi$SR phase, respectively.
In figure \ref{fig:nuB}, we also show the inverse of the correlation length $k_{*}/T = \tilde{\xi}^{-1}$, with the pole of \eqref{eq:corr2}, $k= \pm i k_{*} = \pm i \xi^{-1}$, as a function of $(\tilde{J}-\tilde{J}_{c})$ on a logarithmic scale as approached from the $\chi$SB phase (left) and $\chi$SR phase (right). From a linear fit to the data, we obtain 
\begin{equation}
    \nu_{-} \approx 0.2910, \quad \nu_{+} \approx 0.4980.
\end{equation}
For the $\chi$SR phase, we find that $(\gamma_{+},\nu_{+})$ agree with the mean-field values, $(1,1/2)$. For the $\chi$SB phase, on the other hand, we find that the values of $(\gamma_{-},\nu_{-})$ deviate from those in the Landau theory.
These values $(\gamma_{-},\nu_{-}) \approx (0.58,0.29)$ are the same as in \cite{Matsumoto:2022nqu}, where the quantities are normalized by $B$ and the phase diagram is drawn in the plane of $(T/\sqrt{B},J/B^{3/2})$.
This coincidence is expected because the fixed $B/T^{2}$ line and the fixed $T/\sqrt{B}$ line are equivalent up to a constant factor, and the power of the susceptibility divergence is unchanged. Therefore, the above results are regarded as a consistency check of our earlier results \cite{Matsumoto:2022nqu}.

Secondly, we consider another path to the TCP, such that the ratio of the effective temperature to the heat bath temperature $T_{*}/T$ is fixed to the value at the TCP, which is shown with the dashed line in figure \ref{fig:phase1}. Interestingly, we find that the numerical value of the effective temperature at the TCP nearly coincides with the heat bath temperature: $T_{*}/T \approx 0.9998$. We emphasize that this coincidence is far from trivial since the effective temperature is a complicated function of the external electric and magnetic fields and the background D7-brane configuration as shown in \eqref{eq:Teff}. 
This result indicates that the temperature felt by the fluctuation at the TCP is almost identical to the heat bath temperature.

\begin{figure}
    \centering
    \includegraphics[width=0.95\linewidth]{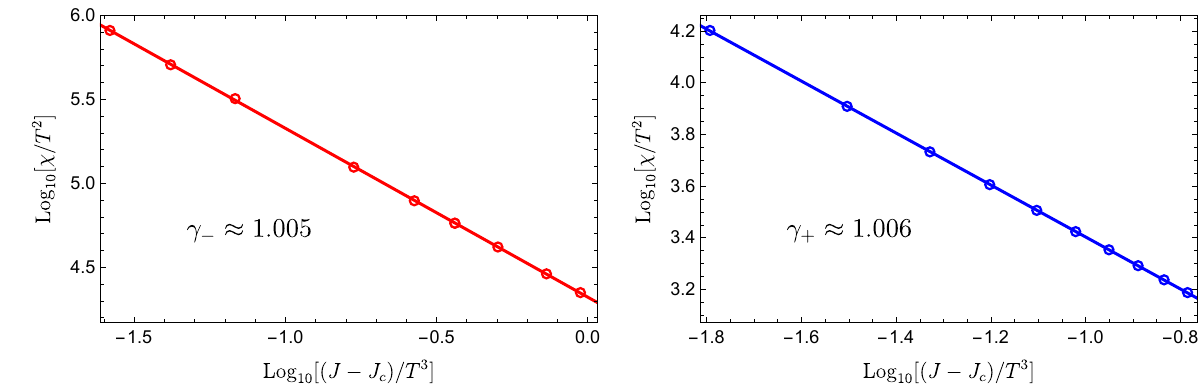}
    \caption{The susceptibility $\chi/T^{2}$ as a function of $(J-J_{c})/T^{3}$ in logarithmic scale when the TCP is approached from the $\chi$SB phase (left) and $\chi$SR phase (right) along the fixed $T_{*}/T$ line. The data points are the numerical results and the solid line denotes the linear fit to them. The corresponding values of $\gamma$ are also shown.}
    \label{fig:gammaT}
\end{figure}

\begin{figure}
    \centering
    \includegraphics[width=0.95\linewidth]{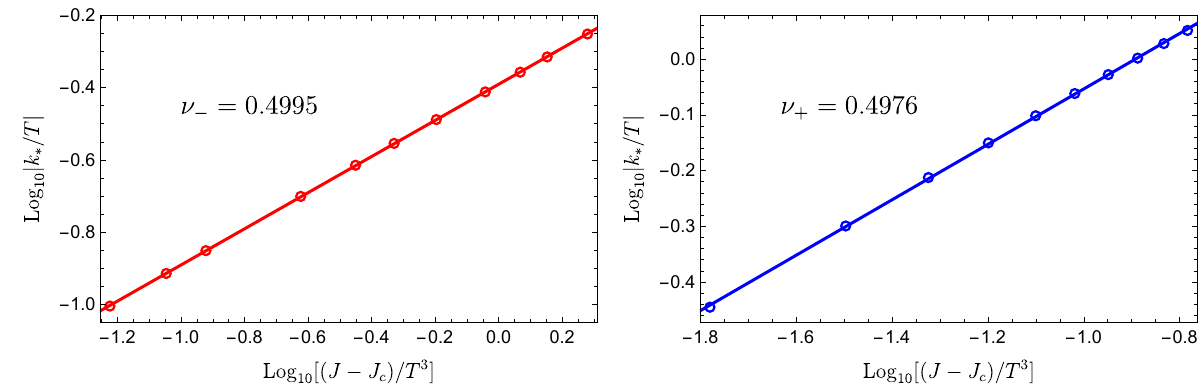}
    \caption{The inverse of the correlation length $k_{*}/T$ as a function of $(J-J_{c})/T^{3}$ in logarithmic scale when the TCP is approached from the $\chi$SB phase (left) and $\chi$SR phase (right) along the fixed $T_{*}/T$ line. The data points are the numerical results and the solid line denotes the linear fit to them. The corresponding values of $\nu$ are also shown.}
    \label{fig:nuT}
\end{figure}

Along the line of fixed $T_{*}/T$, we compute the susceptibility and the correlation length, and determine the values of $(\gamma_{\pm},\nu_{\pm})$. Figure \ref{fig:gammaT} shows the susceptibility and figure \ref{fig:nuT} shows the inverse of the correlation length in the $\chi$SB phase (left) and the $\chi$SR phase (right).
From a linear fit to the data, we obtain
\begin{equation}
    \gamma_{-} \approx 1.005, \quad \gamma_{+} \approx 1.006,
\end{equation}
and 
\begin{equation}
    \nu_{-} \approx 0.4995, \quad \nu_{+} \approx 0.4976.
\end{equation}
Both sets of exponents agree with the mean-field values. This is because the line of fixed $T_{*}/T$ approaches the TCP asymptotically tangential to the second-order phase transition line as shown in figure \ref{fig:phase1}, so that the equilibrium Landau theory reviewed in section \ref{sec:Landau} applies along this path. 

\section{Conclusions and discussion} \label{sec:conclusion}
In this paper, we have investigated the critical phenomena at the current-driven TCP in NESS associated with chiral symmetry breaking in the D3/D7 model in the presence of external electric and magnetic fields.
In particular, we focus on the critical exponents $(\gamma,\nu)$, which are known to deviate from the mean-field values in the $\chi$SB phase \cite{Matsumoto:2022nqu}, and explore whether a Landau-like effective description exists even for this current-driven TCP. Our primary interest is the role played by the effective temperature in these critical phenomena.

We consider two distinct paths to the TCP in the phase diagram with the plane of $(B/T^{2},J/T^{3})$: the line of fixed $B/T^{2}$ and fixed $T_{*}/T$.
In the former case, we reproduce the same values of $(\gamma,\nu)$ as in \cite{Matsumoto:2022nqu}, where the phase diagram is drawn with differently normalized quantities $(T/\sqrt{B},J/B^{3/2})$. This is the expected result since the paths of fixed $B/T^{2}$ and $T/\sqrt{B}$ are identical and hence yield the same power-law divergences of the susceptibility and the correlation length.
In the fixed $T_{*}/T$ case, by contrast, we find that $(\gamma,\nu)$ agree with those of the Landau theory in both the $\chi$SB and $\chi$SR phases. Note that the deviation from the Landau theory occurs only in the $\chi$SB phase: along the non-tangential path, $(\gamma_{+},\nu_{+})$ in the $\chi$SR phase already agree with the mean-field values \cite{Matsumoto:2022nqu}. What we have found here is that this asymmetry between the two phases disappears once the TCP is approached along the line of fixed $T_{*}/T$. This line approaches the TCP asymptotically tangential to the second-order phase transition line in the $\chi$SB phase, whereas non-tangentially in the $\chi$SR phase, as shown in figure \ref{fig:phase1}. 
This suggests that a Landau-like effective theory applies along this path in both the $\chi$SB and $\chi$SR phases, whereas a nontrivial extension of the Landau theory is necessary for a more general description of the current-driven TCP in NESS.

We also numerically find that the effective temperature at the TCP nearly coincides with the heat bath temperature, $T_{*}/T \approx 0.9998$. We have not been able to explain this coincidence analytically because the effective temperature is a complicated function of the external fields and the D7-brane configuration, and the TCP lies in a highly nonlinear region.  
Whether the effective temperature plays an essential role in selecting this tangential direction, or the agreement merely reflects the tangency itself, remains to be clarified.

Several directions remain open. First, it would be important to determine the remaining tricritical exponents, in particular the crossover exponent, and to test the tricritical scaling relations at the current-driven TCP. 
Second, examining whether a similar structure appears in other holographic NESSs would clarify the generality of the present observation. 
Last but not least, a nontrivial extension of the Landau theory for the current-driven TCP should be addressed in future work.
We hope that these directions will shed light on the universal description of critical phenomena far from equilibrium.

\section*{Acknowledgments}
The work of MM is supported by Shanghai Sci-tech Co-research Program (Grant No. 25HB2700600) and the Scholarship Fund for Young Researchers from the Promotion and Mutual Aid Corporation for Private Schools of Japan.
The work of SN is supported in part by JSPS KAKENHI Grant No.~JP25K07174, and the Chuo University Personal Research Grant.

\appendix
\section{Critical phenomena with $B$-normalized quantities} \label{sec:appA}
\begin{figure}
    \centering
    \includegraphics[width=0.8\linewidth]{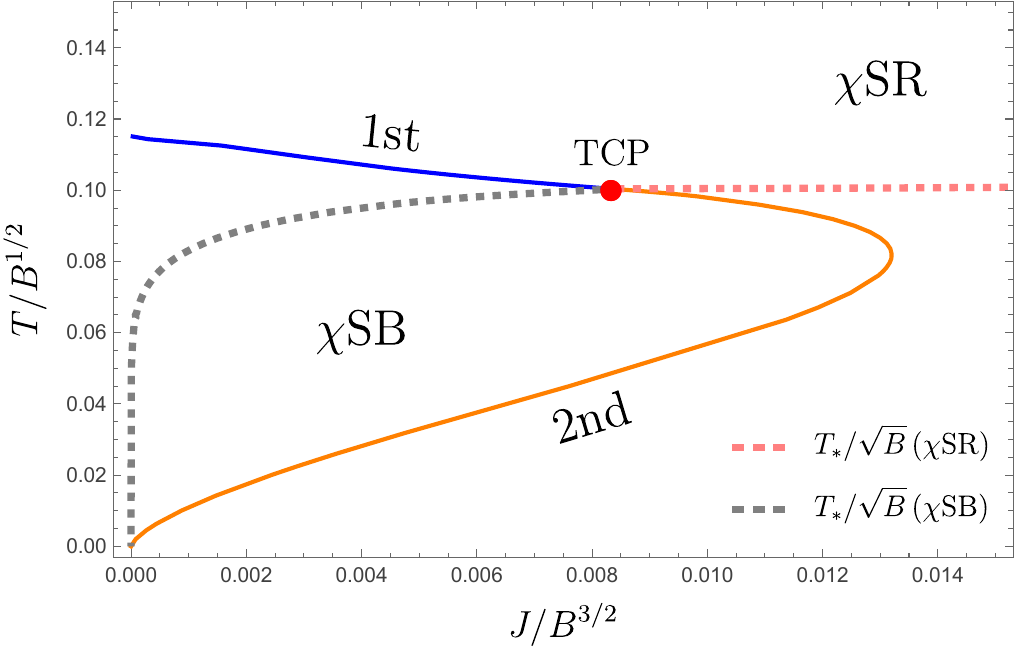}
    \caption{Phase diagram with the plane of $(T/\sqrt{B},J/B^{3/2})$, originally shown in \cite{Matsumoto:2022nqu}. The chiral symmetry-broken ($\chi$SB) phase and chiral symmetry-restored ($\chi$SR)  phase are separated by the first-order phase transition line (blue solid) and the second-order phase transition line (orange solid). 
    The point where the first-order phase transition line changes into the second-order phase transition line is the TCP (red circle).
    The lines for $T_{*}/\sqrt{B}$ fixed at the TCP value for both phases are denoted by the dashed lines. }
    \label{fig:phase2}
\end{figure}

In this appendix, we study the critical exponents $(\gamma,\nu)$ in the phase diagram with the plane of $(T/\sqrt{B},J/B^{3/2})$ as shown in figure \ref{fig:phase2}.
As distinct from the line of fixed $T_{*}/T$ in figure \ref{fig:phase1}, the line of fixed $T_{*}/\sqrt{B}$ approaches the TCP asymptotically non-tangentially to the second-order phase transition line.

\begin{figure}
    \centering
    \includegraphics[width=0.95\linewidth]{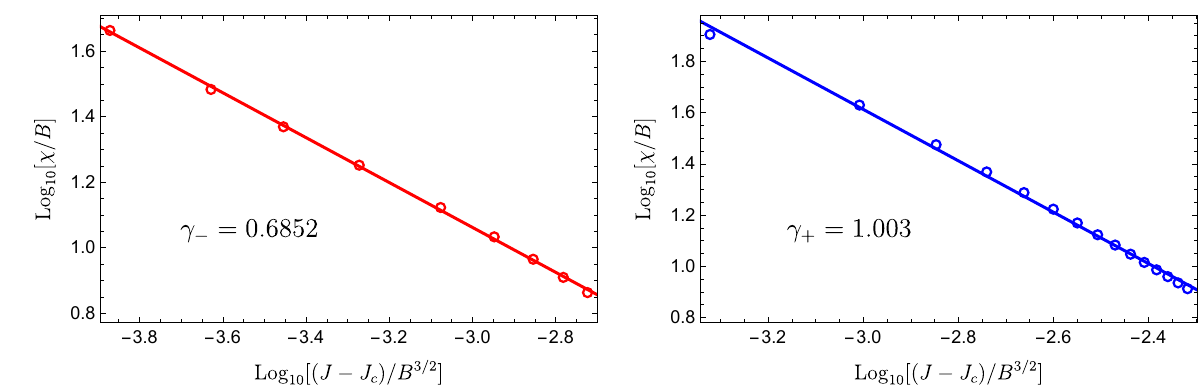}
    \caption{The susceptibility $\chi/B$ as a function of $(J-J_{c})/B^{3/2}$ in logarithmic scale when the TCP is approached from the $\chi$SB phase (left) and $\chi$SR phase (right) along the fixed $T_{*}/\sqrt{B}$ line. The data points are the numerical results and the solid line denotes the linear fit to them. The corresponding values of $\gamma$ are also shown.}
    \label{fig:gammaBB}
\end{figure}

\begin{figure}
    \centering
    \includegraphics[width=0.95\linewidth]{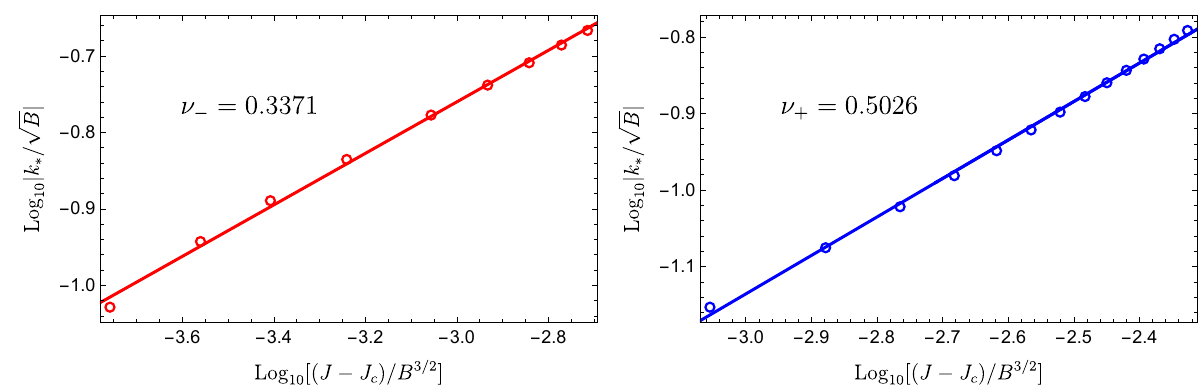}
    \caption{The inverse of the correlation length $k_{*}/\sqrt{B}$ as a function of $(J-J_{c})/B^{3/2}$ in logarithmic scale when the TCP is approached from the $\chi$SB phase (left) and $\chi$SR phase (right) along the fixed $T_{*}/\sqrt{B}$ line. The data points are the numerical results and the solid line denotes the linear fit to them. The corresponding values of $\nu$ are also shown.}
    \label{fig:nuBB}
\end{figure}

We show the results of the susceptibility and the inverse of the correlation length in figure \ref{fig:gammaBB} and figure \ref{fig:nuBB}, respectively.
We define the critical exponents in the same manner as in the main text,
\begin{equation}
    \bar{\chi} \propto \abs{\bar{J}-\bar{J}_{c}}^{-\gamma}, \quad \bar{\xi} \propto \abs{\bar{J}-\bar{J}_{c}}^{-\nu},
\end{equation}
where the quantities with bar are normalized by $B$, such as $\bar{\chi}=\chi/B$, $\bar{\xi} = \xi \sqrt{B}$, and $\bar{J} = J/B^{3/2}$.
Then, we obtain 
\begin{equation}
    \gamma_{-} \approx 0.6852, \quad \gamma_{+} \approx 1.003,
\end{equation}
and 
\begin{equation}
    \nu_{-} \approx 0.3371, \quad \nu_{+} \approx 0.5026.
\end{equation}
We find that $(\gamma_{-},\nu_{-})$ clearly differ from the result along the line of fixed $T/\sqrt{B}$ studied in \cite{Matsumoto:2022nqu}: $(0.58,0.29)$, whereas $(\gamma_{+},\nu_{+})$ agree with the mean-field values. Note that the scaling relation between them still holds, $\gamma = \nu(2-\eta)$ with $\eta=0$. This result indicates that the values of $(\gamma_{-},\nu_{-})$ are not universal for all non-tangential paths, suggesting that the classification of paths at an equilibrium TCP does not straightforwardly hold in the NESS.

\section{Equations for fluctuations} \label{sec:appB}
The quadratic action for the fluctuations, $\tilde{\Phi} =\{\delta \theta, \delta a_{x}\}^{T}$, is given by
\begin{equation}
	S^{(2)} = -\frac{\mathcal{N}}{2}\int \dd[4]{x}\dd{u} R(u)\Big[
		\partial_{\alpha} \tilde{\Phi}^T \tilde{A}^{\alpha\beta} \partial_{\beta} \tilde{\Phi}
  		+ 2 \tilde{\Phi}^T \tilde{B}^{\alpha} \partial_{\alpha} \tilde{\Phi}
  		+ \tilde{\Phi}^T \tilde{C} \tilde{\Phi}
	\Big],
	\label{eq:effective_action_Appendix}
\end{equation}
with coefficient matrices:
\begin{subequations}
\begin{gather}
	\tilde{A}^{\alpha\beta}
	=
	\gamma^{\alpha\beta}P(u) +
	2 \delta^{[\alpha}_{t} \delta^{\beta]}_u Q(u),\\
	\tilde{B}^{\alpha} = 3 \tan\theta
	\begin{bmatrix}
		- M^{\alpha u} \theta'(u) & M^{\alpha x}\\
		0 & 0
	\end{bmatrix},~~~
	\tilde{C} = 
	- 3 (1 - 2\tan^{2} \theta)
	\begin{bmatrix}
		1 & 0\\
		0 & 0
	\end{bmatrix},
\end{gather}
\label{eq:coeffs_position}
\end{subequations}
where
\begin{subequations}
\begin{gather}
	P(u) \equiv
	\begin{bmatrix}
		1- \gamma^{uu}\theta'(u)^{2} & -M^{xu} \theta'(u)\\
		-M^{xu} \theta'(u) & \gamma^{xx}
	\end{bmatrix},\\
	Q(u) \equiv
	(M^{tx} \gamma^{uu} + \gamma^{tu} M^{xu})\theta'(u)
	\begin{bmatrix}
		0 & 1\\
		-1 & 0
	\end{bmatrix},
\end{gather}
\end{subequations}
with $R(u) = - {\cal{L}}_{\rm D7}/{\cal{N}}$.
Here $M^{ab}$ is the inverse matrix of $M_{ab} = g_{ab} + F_{ab}$ and $\gamma^{ab}$ is the inverse matrix of the open string metric defined in \eqref{eq:osm}.
\(P\) is a symmetric matrix, and \(Q\) is an antisymmetric matrix.
\(\tilde{A}^{\alpha\beta}\) satisfies \((\tilde{A}^{\alpha\beta})^T = \tilde{A}^{\beta\alpha}\) by definition.

The equation of motion for $\delta \theta=\vartheta(u)e^{ i k z}$ on the trivial background $\theta=0$ is given by
\begin{equation}
 F \vartheta''
 +\left[\frac{u^3 f h'^2 \left(u f'-6 f\right)}{2} -\frac{F+ 2f}{u}-\frac{f'}{2 f}F \right]\vartheta'
 +\Biggl[ \Biggr. \Bigl(3-k^{2}u^{2} \Bigr)u^{2}h'^{2}+ \frac{\left(3-k^{2}u^{2}\right)}{u^2 f} F \Biggl.\Biggr]\vartheta=0,
 \label{eq:eom0}
\end{equation}
where $F(u)=\left(B^{2}u^{4}+1 \right)f(u)-E^{2}u^{4}$.
Here, $h^{\prime}$ that appears in (\ref{eq:eom0}) is given by 
\begin{equation}
	h'(u)^{2} = -\frac{J^{2}u^{2}F}{\left(J^{2}u^{6}-f\right)f^{2}}.
 \label{eq:hprime}
\end{equation}
The location of the effective horizon is determined by $F(u_{*})=0$.
Note that the current density $J$ is given so that both the numerator and the denominator of (\ref{eq:hprime})  simultaneously become zero at $u=u_{*}$.

On the nontrivial background of $\theta\neq 0$, $\delta\theta$ couples to the fluctuation of the $x$-component of the gauge field $\delta A_{x}=a(u)e^{ i k z}$.
The equations of motion for these perturbations are given by
\begin{eqnarray}
	\vartheta'' + A\vartheta' +Ba' + C\vartheta + D a&=&0, \\
	a'' + \tilde{A}\vartheta' +\tilde{B}a' + \tilde{C}\vartheta + \tilde{D} a&=&0,
\end{eqnarray}
where
\begin{align}
\begin{autobreak}
	A=
    \frac{1}{2uF} \Biggl[ \Biggr. 12ufF\theta' \tan \theta +3u^{2}f^{2}\theta'^{2}\left( uF'-8F\right) +u^{4}f^{2}h'^{2}\left( uf'-6f\right) +F\left(uf'-2f \right) +f\left(uF'-4F \right) \Biggl. \Biggr],
\end{autobreak}
\\
\begin{autobreak}
	\tilde{A}=
	\frac{u f h' \theta '}{F}\left( uF'-8F \right),
\end{autobreak}
\\
\begin{autobreak}
	B=
	\frac{u^2 f h'}{F} \left[ u\theta' \left( u f'-6 f \right)+6 \tan \theta \right],
\end{autobreak}
\\
\begin{autobreak}
	\tilde{B}=
	\frac{1}{2 F u f} \Biggl[ \Biggr. u^{2}f^{2}\theta'^{2} \left( uF'-8F \right) +3u^{3}f^{2}h'^{2}\left( uf'-6f\right) +3F\left(uf'-2f\right) -f\left( uF'-4F\right) \Biggl. \Biggr],
\end{autobreak}
\\
\begin{autobreak}
	C=
	-\frac{k^2}{F f} \Bigl[ u^{4}f^{2}h'^{2}+ F\left(u^{2}f\theta'^{2} +1\right) \Bigr]
	+\frac{3 \sec^{2} \theta}{F u^2 f} \Bigl[ u^{4}f^{2}h'^{2}+ F\left(u^{2}f\theta'^{2} +1\right) \Bigr],
\end{autobreak}
\\
\begin{autobreak}
	\tilde{C} = 
    0,
\end{autobreak}
\\
\begin{autobreak}
	D=
    0,
\end{autobreak}
\\
\begin{autobreak}
	\tilde{D} =
	-\frac{k^2}{F f} \Bigl[ u^{4}f^{2}h'^{2}+ F\left(u^{2}f\theta'^{2} +1\right) \Bigr]
	+\frac{3 \sec^{2} \theta}{F u^2 f} \Bigl[ u^{4}f^{2}h'^{2}+ F\left(u^{2}f\theta'^{2} +1\right) \Bigr].
\end{autobreak}
\end{align}

\bibliography{main}
\bibliographystyle{jhep}

\end{document}